\documentclass{article}
\usepackage{graphicx} 

\usepackage{xurl} 
\usepackage{hyperref}
\usepackage{biblatex}
\usepackage{amsmath}
\usepackage{tabularx}
\usepackage{booktabs}
\usepackage{array}
\usepackage{orcidlink}

\usepackage{tikz}
\usetikzlibrary{arrows.meta,positioning,calc,decorations.pathmorphing,backgrounds}

\definecolor{alphaC}{RGB}{46, 90, 160}   
\definecolor{gammaC}{RGB}{25, 140, 115}   
\definecolor{betaC}{RGB}{114, 76, 160}    
\definecolor{epsC}{RGB}{201, 140, 32}     
\definecolor{deltaC}{RGB}{190, 65, 65}    
\definecolor{pathMain}{RGB}{36, 120, 210} 
\definecolor{pathHalo}{RGB}{120, 180, 245}
\usepackage{amsmath,amssymb,booktabs,array}   
\usepackage{booktabs}          
\usepackage{array}             
\usepackage{tabularx}          
\usepackage{makecell}  

\definecolor{alphaC}{RGB}{46, 90, 160}   
\definecolor{gammaC}{RGB}{25, 140, 115}  
\definecolor{betaC}{RGB}{114, 76, 160}   
\definecolor{epsC}{RGB}{201, 140, 32}    
\definecolor{deltaC}{RGB}{190, 65, 65}   

\definecolor{pathA}{RGB}{36,120,210}     
\definecolor{pathB}{RGB}{230,145,40}     
\definecolor{pathC}{RGB}{160,70,190}     

\definecolor{haloA}{RGB}{120,180,245}
\definecolor{haloB}{RGB}{255,200,120}
\definecolor{haloC}{RGB}{210,160,235}

\tikzset{
  labelbox/.style={
    rounded corners=2pt, fill=white, draw=black!20,
    inner sep=2pt, font=\scriptsize, align=center
  }
}

\author{
  Angjelin Hila \orcidlink{0009-0005-0481-0443} \\
  School of Information, University of Texas at Austin \\
  \texttt{ahila@utexas.edu}
}

\title{The Human-AI Delegation-Verification Dilemma: Individual Strategies, Collective Equilibria and Sociotechnical Lock-in}

\begin{document}

\maketitle

\begin{abstract}
This paper takes an ecological approach toward large-scale models of hybrid human-AI intelligence. Emerging models of human-AI interaction predominantly advance the complementarity thesis variously dubbed human-AI collaboration and human-AI hybrid intelligence. However, this constitutes an over-simplification of the modalities of human-AI interaction and possibility-space for both individual and collective action that human-AI interaction potentiates. To fill these gaps, this paper develops a decision and game-theoretic approach to the human-AI delegation-verification dilemma. First, we map out canonical decision-theoretic strategies that account for adaptive user trajectories, modeling how agents transition between strategies based on interaction feedback to reach stable equilibria. Second, we scale individually stable strategies to collective equilibria using three extrapolation principles: (a) non-communicative aggregation (b) local social signaling and (c) institutional norms setting. The analysis identifies the emergence of sociotechnical lock-in, a macro-behavioral state where individually adaptive delegation, in the absence of communicative and institutional safeguards, aggregates into a systemic collective action problem modeled as a prisoner’s dilemma that degrades shared epistemic standards. We argue that adoption under higher communicative standards and institutional norms can mitigate suboptimal collective equilibria by imposing social commitments on individual users.
\end{abstract}

\providecommand{\keywords}[1]
{
  \small	
  \textbf{\textit{Keywords---}} #1
}


\section{Introduction}
In his celebrated \textit{The Sciences of the Artificial} \cite{simon1996} Herbert Simon defined artificial intelligence as “systems designed by humans in order to attain goals, and function”. Four decades later, the philosopher of AI, Nick Bostrom predicted the capacity for self-optimization to be the juncture at which artificial intelligence will begin to exponentially eclipse human intelligence \cite{bostrom2016}. In the last half-decade, the emergence of generative AI (GenAI) in the form of large language models (LLMs) in language generation and stable diffusion models (SD) in image generation have renewed discourse surrounding the relationship between human and AI agencies. In light of these developments, human-AI collaboration and human-AI hybrid intelligence have emerged as dominant theoretical frameworks. However, consensus regarding the precise framing of these dynamics remains elusive and the terms of collaboration and hybrid intelligence fluid. 

Proponents of the hybrid intelligence thesis espouse positions that tend to equivocate on the ontological status of current AI models between (a) tool-like extensions of human plans and (b) cognitive agents in their own right. In the former camp, hybrid intelligence theorists implicitly conceptualize hybrid intelligence within a version of extended-mind theory, which posits that human cognition extends into environmental tools \cite{clark_chalmers_1998}. The extended mind hypothesis posits functional requirements that delineate which external entities partake in cognitive coupling and which lie peripheral to it \cite{clark_chalmers_1998, clark_2008, clark_2013_whatever_next, menary_2010}.  The latter camp, on the other hand, recognizes that human-AI coexistence can evolve toward a competitive and non-mutualistic dynamic \cite{Peeters2021, Noller2025, Shkurko2025, Alsobay2025, Zhu2025}. The persisting ambiguity between these two opposing designations in conjunction with the lack of ecological explorations of the evolving shape of human-AI interaction provides the impetus for this paper. 

In this paper, we develop a theoretical framework for understanding the looping of AI systems with linguistic and decision-making capabilities into large-scale frameworks of cooperation and competition. Specifically, we develop a model of human-AI interaction that leverages the formal apparatus of decision theory and evolutionary game theory. In both cases, we model human-AI interaction as an evolutionary, path-dependent process, accounting for both the individual decision-theoretic trajectories and the collective population dynamics that emerge through repeated, multi-shot interactions. We highlight several stable evolutionary strategies as possible collective equilibria that encompass both optimal and suboptimal configurations at both the individual and collective scales. In order to do so, we first formalize a decision-theoretic space of possible human-AI strategies. Subsequently, we extrapolate individual decision-theoretic strategies into multi-agent scenarios. In particular, we show how individual theoretic strategies scale into cooperation and coordination dynamics where costs and benefits are contingent on (a) individual interaction dynamics and (b) emergent population patterns shaped by signaling, communication, and institutional standards. Our analysis suggests that human-AI interaction should not be ipso-facto framed as a collaborative game but rather as a combination of (a) subjective utility maximization and (b) a social game subject to normative and collective coordination controls. This framing allows for human-AI communication to exhibit both zero-sum and non-zero sum outcomes, and identify factors that scale to optimal and suboptimal collective equilibria.

\section{Background}

The emergence of human-LLM interaction and the increasing integration of LLMs into human workflows have engendered theories that frame the human-AI relationship as fundamentally collaborative \cite{Cabreraetal2023, fragiadakis2024evaluating, PuertaBeldarrain2025}. According to the collaborative thesis, human-AI interaction fundamentally yields a symbiosis that in broad terms augments human agency \cite{jarrahi_lutz_newlands_2022}. To this effect, research within human-AI collaboration is primarily oriented toward the alignment of human and AI agencies with the aim of seamless communicative articulation \cite{Cabreraetal2023, fragiadakis2024evaluating, Gomezetal2024, Fragiadakis2024}. For Wang et al \cite{wang_churchill_maes_fan_shneiderman_shi_wang_2020} ``collaboration involves mutual goal understanding, preemptive task co-management and shared progress tracking" \cite{wang_churchill_maes_fan_shneiderman_shi_wang_2020}. In line with these aspirations, Jarrahi et al \cite{jarrahi_lutz_newlands_2022} have postulated the emergence of synergist intelligence between human and AI agencies. Their synergist thesis results from the observation that ``devising autonomous systems is almost impossible for many real-world scenarios where humans need to “stay in the loop” to maintain the sociotechnical system's versatility and adaptability in relation to new tasks and environments" \cite{jarrahi_lutz_newlands_2022}. According to Jarrahi et al \cite{jarrahi_lutz_newlands_2022} mutual gaps human and AI tacit knowledge impedes seamless human-AI symbiosis. On the other hand, emergent lines of research resist the symbiotic framing by characterizing the human-AI relationship in labor exploitative terms \cite{Sarkar2023, Earp2025, Wei2024}. These tensions within human-AI interaction underscore a deeper indeterminacy with respect to the kinds of payoffs human-AI collaboration theoretically promises and the kinds of quality and efficiency trade-offs that concrete interaction presently produces. In consequence, we seek to close these gaps in understanding by mapping human-AI interaction from an ecological perspective that integrates the theoretical apparatus of decision theory and evolutionary game theory. Accordingly, we situate our approach first in the theory of human-technology relations and subsequently in models of collective coordination and action. 

The symbiosis thesis finds its obverse in the reverse-adaption thesis. Two theoretical senses of reverse adaption bear theoretical relevance to our analysis of human-AI relations: (a) Langdon Winner's sense as the organization of human ends to suit technological means \cite{winner1978autonomous} and (b) Andy Clark's sense as environmental adaptations that exploit human behaviors for extraneous ends\cite{clark2003naturalborncyborgs}. In human-technology relations, both notions of reverse adaptation describe situations in which technological infrastructure molds human ends to fit the system's available means. In the context of human-technology relations, Winner defines reverse adaptation as ``the adjustment of human ends to match the character of the available means" \cite{winner1978autonomous}. Clark explicitly describes a relationship in which the environment exploits human reward mechanisms to produce suboptimal agent behaviors \cite{clark_2008}. Drawing on Clark, Heath argues that the resources for reverse adaption available to organized actors in the cultural sphere create a hostile environment for individual rational decision-making \cite{heath2014enlightenment}. Both Clark and Heath recognize that interactive scenarios of predation and parasitism in addition to mutualism occur in the cultural sphere between rational and sub-rational actors. In the private sphere, it is often the case that individual interests are at odds with corporate interests. Heath argues that conflicts of interest between individual human agents and corporate agents stack in favor of the latter because the rational decision-making resources of organizations dwarf the rational deliberation resources of individuals \cite{heath2014enlightenment}. Examples of such phenomena include advertising, junk food, and social media apps, to name a few, where product design has adapted to exploit evolutionary and cognitive heuristics to the benefit of the producer and often the detriment of the consumer \cite{heath2014enlightenment}. Therefore, what in theoretical economics are framed as exchanges of mutual benefit where both parties increase their utility often evolve into predatory, zero-sum relationships where an increase in utility for the corporation results in a decrease in utility to the consumer. These dynamics are a fortiori replicated in human-AI relations in part because, as Winner acknowledges, consumers are passive receivers of a system they did not design and, consequently, must adapt their ends to the means inscribed in the system \cite{winner1978autonomous}. With respect to LLM chatbots specifically, these exploitative relations are compounded by the fact that chatbots can be engineered to foster emotional connection and dependency in human users. Reverse adaption, therefore, underscores one dimension of the decision-theoretic landscape that besets human-AI interaction. It does not, however, preclude the possibility that individual human agents can utilize AI to increase personal utility and the efficiency of personal effectance, the ability to achieve goals and produce intended effects in the environment. 

Beyond individual choice, a dimension of research relevant to our goals in this paper concerns the study of coordination dynamics and collective behavior. Two seminal works in this area constitute \textit{The Logic of Collective Action} by Mancur Olson and\textit{ Governing the Commons: The Evolution of Institutions for Collective Action} by Elinor Ostrom. Olson challenged the then assumption that organizations function to further common collective interests and argued that members of large organizations have a rational incentive to defect, called the free-rider problem \cite{olson1965logic}. Given that all groups or organizations dole out some public or collective good to their members, Olson argued that incentive to contribute is greatest when the collective benefit accrued exceeds the personal cost of contribution \cite{olson1965logic}. Therefore, because the share of the collective good decreases in proportion to the size of the organization, the incentive for rational actors is to forego their contribution while still reaping the collective benefit, also known as ``free riding" \cite{olson1965logic}. Olson reasoned that because rational incentive to ``free-ride" increases in proportion to the size of an organization, (a) smaller groups or organizations are more effective in coordinating their interests and (b) large organizations must resort to some coercive arrangement to levy contributions \cite{olson1965logic}. As such, Olson underscored a major problem beleaguering collective action and group coordination. Framed as an n-person or generalized prisoner's dilemma, however, where the costs of defection are contingent on the epistemic states of agents, perfectly rational agents in the game-theoretic sense always produce suboptimal outcomes \cite{axelrod2006evolution}. In particular, one-shot prisoner's dilemmas suffer from the problem of epistemic opacity, where agents have no knowledge of other agents' intentions. Simulating computer program tournaments over iterated interactions, Axelrod demonstrated that repeated or many-shot prisoner's dilemmas (RPD) converged on tit-for-tat over defection as an evolutionary stable strategy (ESS) \cite{axelrod2006evolution}. Building on Axelrod's findings, Elinor Ostrom, critic of Olson and rational choice theory, showed that Olson's idealized scenarios belie the empirically observed dynamics of cooperation where payoffs depend on repeated interactions that produce group stratification through communication incentivizing recriprocators over rational maximizers \cite{ostrom1990governing}. In our analysis we will pursue possible evolutionary stable strategies that can emerge from scaled human-AI interaction. 

Another facet of our collective coordination approach to human-AI interaction concerns the dynamics of technological adoption, critical-mass phenomena, and the constraints on individual choice that emergent global patterns impose. In his seminal work \textit{Micromotives and Macrobehaviors}, Thomas Schelling argued that collective coordination patterns emerge from local individual choices that he called microbehaviors. According to Schelling, the microbehaviors of individual actors obey subjective critical threshold criteria that in aggregate produce collective critical-mass behaviors \cite{schelling1978micromotives}. Schelling called individual critical thresholds k-values, and argued that their population variance determines the compounding patterns of macrobehaviors \cite{schelling1978micromotives}. To illustrate, consider a party where no one has started dancing. If a brave soul becomes the first to do so, they risk looking foolish, which means they have the lowest threshold for participation. At the same time, the participation of one person suffices for some to join in, whereas for others, a group of a few activates their threshold. In effect, the lower-threshold actors instigate the next-threshold actors in the distribution and so on, producing a compounding effect where the majority join the dance. In addition to activation thresholds, agents may also assign subjective weights to other agents, further influencing thresholding behavior. Broadly, Schelling observed that across a variety of critical-mass phenomena, cascading effects depend on the distribution of agent policies. Gronovetter extended Schelling's insights by modeling variable policies that absorb varying degrees of environmental influence \cite{granovetter1978threshold}. In the context of human-AI interaction we will adopt the assumption of non-ergodicity, where initial conditions do not predict outcomes, and the view that agent policies are dynamic. 

Extending Schelling and Gronovetter's insights, Brian Arthur argued in \textit{Competing Technologies, Increasing Returns and Lock-in by Historical Events} that technological adoption patterns evolve toward a state of inflexibility or ``lock-in" marked by increasing  rather than diminishing returns, contravening the predictions of orthodox economics \cite{arthur1989competing}. Arthur showed that the utility of choosing an inferior product when it has reached mass adoption is greater than that of a superior alternative due to learning effects, network externalities, and economies of scale \cite{arthur1983generalized, arthur1989competing}. Taking an evolutionary game theory approach, Skyrms showed that lock-in phenomena emerge broadly in social games where replication dynamics that select strategies that outperform the population average filter through symmetry breaking phenomena such as agent assortativity and the emergence of conventions from communication and signaling \cite{skyrms1996evolution}. Whereas Arthur framed lock-in as a cause of market inefficiencies due to increasing returns, we will argue that sociological lock-in produces collectively suboptimal self-reinforcing norms. We will extend Arthur's lock-in thesis by arguing that a multi-factor model that takes into account several causal factors of lock-in shows path dependencies that ultimately settle into suboptimal equilibria.

\section{Game-Theoretic Interaction Space}
Approaching human-AI interaction from a decision and game-theoretic perspective requires first addressing the human-AI cognitive asymmetry. Whereas traditional decision theory theorizes instrumental reason at the level of individual agents, game theory extrapolates to multi-agent interactions \cite{heath2009communicative}. Classical game-theory entails what I call a principle of indifference about agents by implicitly assigning a belief-desire psychology to all agents. The uniform distribution of a belief-desire psychology across agents produces what I call symmetrical interactions, where agents share a similar cognitive system. Because this assumption cannot be extended to human-AI interaction, a decision and game-theoretic approach to human-AI interaction should take into serious consideration the differences in cognitive constitution and resources between human and AI agents. As we will demonstrate, these differences carry deep implications for theorizing the interaction space both at the individual human-AI interaction scale and scaling the effects to collective behavior. 

A major complication for classical game theory stems from extending the belief-desire psychology to multiple agents. Classical game theory proceeds from a parsimonious ontology of the following triple: (a) states, (b) actions and (c) outcomes. States describe a variable configuration of the world, actions represent available choices given some variable state, and outcomes the effects of actions \cite{heath2009communicative, heath2011}. More specifically, states denote states of the world outside the agent's control, such as whether an AI will hallucinate or not. Actions denote functions from a state to an outcome \cite{heath2009communicative, heath2011}. This formulation follows the Savage triple, which generalized the Von Neuman-Morgenstern utility-maximization procedure from a function over known  external world odds to Bayesian subjective states as subjective utility maximization \cite{savage1954foundations}. Because in a two-agent world, outcomes are determined jointly by the actions of both agents, the actions of one agent will be states for the other and vice versa \cite{heath2011}. As Heath points out, in order to assign probabilities to states, agents must determine the probabilities of each others' actions, which requires estimating their beliefs \cite{heath2011}. Since actions are states for each of their utility functions, anticipating another agent's states entails anticipating their anticipations of one's actions, which leads to an infinite regress of anticipations \cite{heath2011}. Nash equilibria supply solutions to such regress problems by enabling players to select mixed strategies that randomize over all possible combinations \cite{heath2011}. A Nash equilibrium defines a selection of strategies by agents where no agent can improve their payoffs by selecting an alternative strategy while other players' strategies remain equal or ceteris paribus \cite{watson2013strategy}. However, a major problem with Nash equilibria is that they supply mathematical, ad-hoc solutions to decision-selection problems with no pragmatic bearing on actual real-world scenarios. Where multiple possible equilibria exist, therefore, the \textit{equilibrium selection problem} poses a major explanatory challenge for mathematical game theory for how coordinated-action emerges in the face of strategic indeterminacy. 

Because the traditional decision-theoretic model breaks down in social-choice scenarios, evolutionary game theory emerged as an alternative framework for modeling and explaining collective equilibria. Smith and Price showed that evolutionary stable strategies (ESS) emerge as stable proportions of population strategies \cite{maynardsmith1972game, maynardsmith1973logic}. Where Nash equilibria resort to mixed agent strategies, Smith and Price transposed the mixture to a distribution of agent types in the population. This population extension showed that evolutionary stable strategies are a proper subset of Nash equilibria \cite{maynardsmith1982evolution}, while also delimiting the interaction space into possible and impossible evolutionary stable strategies \cite{maynardsmith1982evolution}. Smith and Price defined evolutionary stability as an invader-proof distribution where a random mutation cannot do better against the incumbent than it does against itself and if the incumbent and the mutation are equal, the incumbent must do better against the mutation than the mutation does against itself \cite{maynardsmith1973logic, maynardsmith1982evolution}. These stipulated dynamics ensure the stability of the equilibrium against novel strategies. Taylor et al \cite{taylor1978evolutionary} extended Smith and Price's model from static frequency-dependent payoffs to a replicator evolutionary dynamic where a strategy grows if its payoff exceeds the population average \cite{taylor1978evolutionary}. The replicator dynamics showed that randomized interactions across infinite time stabilize into a mixed-strategy population with equally distributed payoffs \cite{taylor1978evolutionary}. However, real-world dynamics belie the assumptions built into these early models in which agent types are fixed and interactions are randomized. The introduction of mimicry, learning, assortativity, and communication into the mix produces what Skyrms calls symmetry-breaking effects, where path-dependent stable arrangements emerge \cite{skyrms1996evolution}. In finite populations, therefore, where agents can mimic each other's strategies, select the agents with whom they interact based on available information, learn new behaviors and coordinate actions through communication, equilibria are not immutable terminal states but metastable basins liable to continual phase shifts from tipping points caused by changing environmental conditions. 

Consequently, given the developments within game theory, communicative action, and sociology in the last few decades, we theorize equilibria not as static terminal states, as theorized by Nash, but as satisficing attractors that are locally stable but entrained in recurrent dynamics and can always transition toward new stable states when key parameters exceed certain value thresholds. This approach aligns with the assumptions of evolutionary game theory. To give an example, a user's trust may be attenuated when the model hallucinates or repeatedly yields outcomes laced with errors. However, if a greater proportion of outputs satisfy user accuracy criteria, the users will stay in some interactive equilibrium. On the other hand, if the user comes to believe that the model is unreliable in one task domain or wholly unreliable, it will prompt a strategy phase shift leading the user to change strategy and settle into a new satisficing regime. One of the core vectors of human-AI interaction we will address concerns the information asymmetry and differential epistemic states between human and AI agents. Epistemic initial conditions on both agentic sides carry implications for the possible equilibria that can and are likely to obtain in repeated interactions. First, we define agent-AI interaction strategies from uninformative priors to metastable equilibria. Subsequently, we explore the interactive dynamics between human and AI agents within the framework of signaling and communicative games.

\subsection{Decision-Theoretic Strategies}
In this section, we formalize individual human agent strategies within human-AI interaction. Because decision theory concerns the rationality of individual agents, in this section we consider the decision-theoretic choices of human agents with respect to AI in isolation from other agents. We later scaffold to multi-agent dynamics in the game-theoretic interactions section, where our discussion moves to coordination problems and collective equilibria. 

Following subjective utility maximization and bounded rationality \cite{simon1996}, we assume that users will satisfy their individual subjective utility conditions under partial or incomplete information. While the variety of user types is likely to be large, we derive a typology on the basis of the types of subjective utility users are likely to maximize. We classify these into: (a) instrumental efficiency, (b) epistemic assurance, (c) reflective understanding, (d) generative payoff and (d) the prudential mixture of all the above. Users oriented toward maximizing instrumental efficiency will seek to maximize usable output per unit cost of effort expended. Users oriented toward epistemic assurance will seek to maximize the reliability, defensibility, and error control relative to error risk. Users oriented toward reflective understanding will seek to maximize internally reconstructable representations of reasons, relations, and justifications relative to passive acceptance in units of external substitution. Users oriented toward generative payoff will seek to maximize the expansion of their action space through productive variation, novelty, and iterative refinement relative to the path of exploration. Finally, prudential mixture users will seek to maximize a context-sensitive distribution of all the above criteria relative to anticipated stakes, task demands and expected loss from resource misallocation. 

\begin{figure}[t!]
\centering
\includegraphics[width=\linewidth]{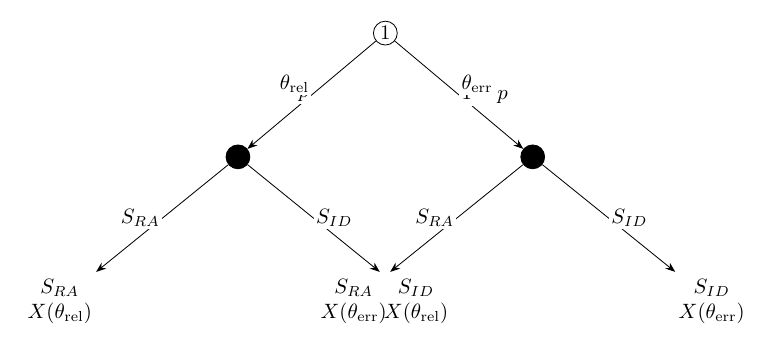}
\caption{Decision diagram illustrating the two strategy poles: Reflective Augmentation and Instrumental Delegation. Nature draws AI reliability (\(\theta\)) and the human agent selects a strategy leading to an outcome space.}
\label{fig:decision_diagram}
\end{figure}

These broad dimensions of utility evaluation correspond to user behavioral propensities that produce action patterns of human-AI interaction. User behavioral propensities form stable basins of attraction based on an internal modeling and prioritization of subjective goods. However, we do not operationalize users as fixed types but as adaptive agents that can recalibrate their priorities and transition between strategies. Specifically, we operationalize users as agents with stable dispositions and sedimented action-patterns yet variable strategy orientations. In this regard, while users will gravitate toward certain strategies more than others, model feedback and complex environmental signals can produce dynamic transitions between strategies.  We consequently sort users by their primary decision orientation in human-AI interaction scenarios. Empirically, user decision-orientations are likely to be biased by an informative prior of AI expectations. In our scenario, we abstract away knowledge of AI expectations in order to constrain the human-AI evolutionary trajectories to the interaction pattern, whence expectations emerge from ongoing feedback. As such, while users possess relatively stable action-pattern histories and partially locked-in dispositions, strategies are variable policy-states that exist independently of particular users and can be entered, exited, and recombined under changing feedback signals. See table ~\ref{tab: strat_definitions} for concise definitions of each unique strategy and table ~\ref{tab: strat_reliability} for a theoretical breakdown of the reliability parameter. Figure ~\ref{fig:decision_diagram} depicts a decision diagram that illustrates the transition from nature's determination of the AI reliability state to the human agent choice of strategy and their consequences, operationalized for simplicity into binary outcomes of reliable and error-laden outputs.

\begin{table}[h!]
\centering
\small
\renewcommand{\arraystretch}{1.25}
\resizebox{\textwidth}{!}{%
\begin{tabular}{>{\raggedright\arraybackslash}p{1.6cm} >{\raggedright\arraybackslash}p{12.0cm}}
\toprule
\textbf{Strategy} & \textbf{Definition} \\
\midrule
$\alpha$ \textit{Reflective Augmentation} &
Policy of adversarial use: treat AI outputs as prompts for scrutiny (objections, counterexamples, alternative derivations), and adopt conclusions only after the human reconstructs and endorses the justification internally. \\

$\gamma$ \textit{Collaborative Synthesis} &
Policy of iterative co-creation: the human provides goals/values/constraints; the AI provides generative search and variation; intermediate outputs are treated as negotiable hypotheses refined through successive turns (a stable interaction mode whose realized equilibrium varies by task parameters). \\

$\beta$ \textit{Supervisory Verification} &
Policy of conditional adoption: the AI generates candidate outputs, and the human performs an explicit audit (fact-checking, consistency tests, source validation, red-teaming) before use; acceptance is contingent on passing verification. \\

$\epsilon$ \textit{Selective Offloading} &
Meta-policy of routing: allocate cognitive effort by stakes, switching among $\alpha$, $\beta$, $\gamma$, and $\delta$ as a function of anticipated loss from error; low-stakes tasks are offloaded, high-stakes tasks are handled with deeper reflection or auditing (often observed as a stable day-level equilibrium). \\

$\delta$ \textit{Instrumental Delegation} &
Policy of default acceptance: treat AI output as the primary recommendation and minimize scrutiny; adopt results with little or no independent reconstruction or auditing, except when errors become salient. \\
\bottomrule
\end{tabular}
}
\caption{Ordinal strategy/policy set from highest internalist/reflective investment ($\alpha$) to maximal instrumental delegation ($\delta$).}
\label{tab: strat_definitions}
\end{table}

\begin{table}[h]
\centering
\small
\renewcommand{\arraystretch}{1.3} 
\begin{tabularx}{\linewidth}{>{\centering\arraybackslash}m{1.2cm} >{\centering\arraybackslash}X >{\centering\arraybackslash}X}
\toprule
\textbf{Code} & $\theta_{\mathrm{rel}}$ (reliable) & $\theta_{\mathrm{err}}$ (error-prone) \\
\midrule
$\alpha$ & $X(\alpha,\theta_{\mathrm{rel}})\subseteq \mathbb{R}^n$ & $X(\alpha,\theta_{\mathrm{err}})\subseteq \mathbb{R}^n$ \\
$\beta$  & $X(\beta,\theta_{\mathrm{rel}})\subseteq \mathbb{R}^n$  & $X(\beta,\theta_{\mathrm{err}})\subseteq \mathbb{R}^n$  \\
$\gamma$ & $X(\gamma,\theta_{\mathrm{rel}})\subseteq \mathbb{R}^n$ & $X(\gamma,\theta_{\mathrm{err}})\subseteq \mathbb{R}^n$ \\
$\delta$ & $X(\delta,\theta_{\mathrm{rel}})\subseteq \mathbb{R}^n$ & $X(\delta,\theta_{\mathrm{err}})\subseteq \mathbb{R}^n$ \\
$\epsilon$& $X(\epsilon,\theta_{\mathrm{rel}})\subseteq \mathbb{R}^n$& $X(\epsilon,\theta_{\mathrm{err}})\subseteq \mathbb{R}^n$\\
\bottomrule
\end{tabularx}
\caption{Outcome correspondence by strategy and AI-reliability state. Each cell is a set of $n$-dimensional consequence vectors.}
\label{tab: strat_reliability}
\end{table}

We now proceed to formalize probabilistic transition dynamics from initial strategies to stable evolutionary basins based on a three variable feedback vector that takes into account variable user trust, error signals, and calculated cognitive effort expended for desired outputs. This is meant to be a schematic representation of a continuous process that is deeply contingent and shaped by the user's subjective experience.  We take the strategy set $S$ we have previously defined as input and define $z$ at $t$ as the signal input that determines the user's evolutionary trajectory per time-step and \(s_t\) the users' current policy state:

\[
S=\{\alpha,\gamma,\beta,\epsilon,\delta\},
\qquad s_t \in S
\]
where \(s_t\) is the user's policy-state (strategy) at interaction episode \(t\). We define the transition vector \(z_t\) as a triple of three core variables that each LLM input embodies: (a) trust or perceived model competence, (b) detected error, and (c) perceived verification burden as cognitive time cost. While in practice these constitute highly correlated variables, they can in principle vary independently, which justifies operationalizing them as distinct values in the signal vector: 

\[
z_t = (\tau_t,\varepsilon_t,\kappa_t)\in \mathbb{R}^3
\]
where \(\tau_t\) denotes trust or perceived competence, \(\varepsilon_t\) denotes error salience, and \(\kappa_t\) denotes perceived verification burden or cognitive cost at time \(t\). 

Taking the interaction invariant axes as epistemic state and repeated interactions, we theorize that user interaction with an AI undergoes an evolutionary trajectory from its initial state, where initial states vary across users and user dispositions, toward some stable interaction equilibrium that satisfies a human agent's subjective satisficing criteria. For this reason, we posit that evolutionary trajectories can vary significantly between users and settle into divergent equilibria. One potential method of resolving these evolutionary trajectories would be through a combinatorial exhaustion from a possible initial strategy to an equilibrium strategy over repeated interactions given some transition rule. 
Below, we show the schematics of these transition dynamics meant to highlight possible evolutionary trajectories and emergent equilibria. However, because our aim is for empirical plausibility and not mathematical exhaustion, we will highlight three types of evolutionary trajectories we deem plausible given prior information about broad categories of user types. Broadly, we categorize users into productivity-oriented or throughput-oriented users and assurance or validity-oriented users. We think that a typology can help sketch plausible evolutionary trajectories that represent emergent individual equilibria. As such, we posit that a user's strategy evolves as a history-dependent transition process:
\[
s_{t+1} \sim T\!\left(\,\cdot \mid s_t,z_t,h_t,\omega_i\right),
\]
where \(s_t\) denotes the current strategy, \(z_t\) the signal vector, \(h_t\) denotes interaction history and \(\omega_i\) denotes user-type parameters, reflectively-oriented, throughput-oriented, assurance-oriented dispositions. \(\sim T_t\) stands for the transition rule, namely the probability of transition from one strategy to another given current input parameters. Equivalently, the dynamics can be expressed in deterministic threshold form, where the transition from one strategy to another given input parameters follows the transition rule $f$ :
\[
s_{t+1}=f(s_t,z_t,h_t,\omega_i).
\]

Given the above schematic formalization, we highlight three evolutionary trajectories and emergent individual user equilibria we believe to be empirically plausible: (a) adaptive recalibration, (b) throughput lock-in and (c) mixed-assurance.  

\begin{figure}[t!]
\centering
\includegraphics[width=\linewidth]{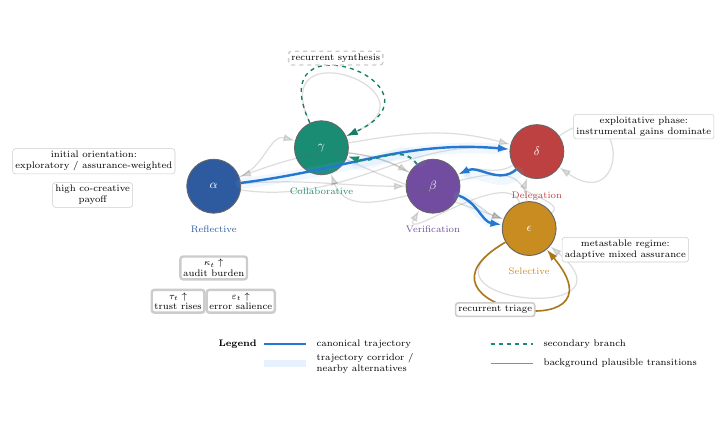}
\caption{Sample strategy-evolution state-space diagram for one canonical repeated-interaction scenario. Colored nodes denote policy-states (\(\alpha,\gamma,\beta,\epsilon,\delta\)); curved arrows denote transitions. The translucent corridor around the main path visualizes nearby plausible pathways (a stylized “least-action” channel) rather than a single deterministic trajectory.}
\label{fig:path_a}
\end{figure}

\paragraph{Adaptive Recalibration}
The adaptive recalibration equilibrium emerges in users with an epistemic orientation toward augmenting their knowledge and understanding rather than maximizing their productivity. As a result, these users are likely to filter LLM outputs through high-intensity reflective evaluation, correcting them where they find discrepancies, errors and lack of nuance, but exploiting the open domain question answering (ODQA) dynamic that LLMs enable to maximum effect. Across repeated interactions, and depending on the sustained quality of LLM outputs, these users are likely to shift into a higher delegation equilibrium as trust rises, but correct this new dynamic once error-patterns emerge and the limits of delegation become apparent. As such, these users are likely to settle into an adaptive metastable equilibrium that is reflectively oriented, delegates where safe, and verifies when stakes are high. Figure ~\ref{fig:path_a} depicts a graphical representation of the transition path from initial strategy to stable equilibrium.  

\begin{figure}[t!]
\centering
\includegraphics[width=\linewidth]{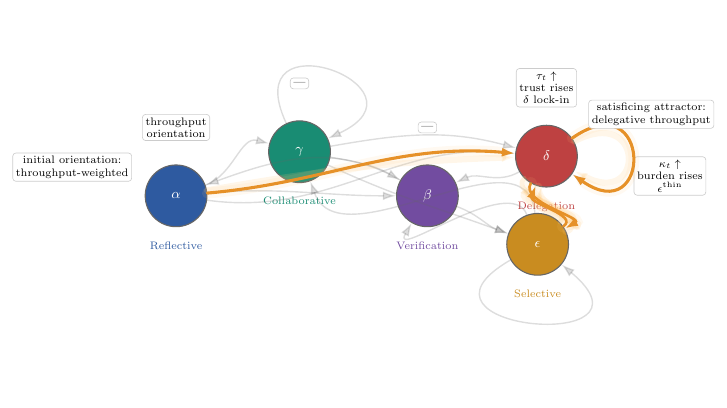}
\caption{Canonical Path B (suboptimal): throughput-weighted users converge to \(\delta\), with thin triage \(\epsilon^{\text{thin}}\) as a fragile mitigation that often reverts under least-effort pressure.}
\label{fig:path_b}
\end{figure}

\paragraph{Throughput Lock-in}
The throughput lock-in equilibrium emerges in users with a tendency of high trust and orientation toward maximizing the quantity rather than the quality of outputs. In other words, these users are throughput oriented. Because of this initial orientation, these users are likely to maximize the delegative potential that human-LLM interaction presents. Due to their high-trust, early-adopter orientation, they are likely to eschew error oversight, opting instead for the trustworthiness cues embodied in the confidence and fluency of model outputs. Consequently, these users are likely to delay the discovery of masked errors or weakly sanction them when necessary to obtain merely functional outcomes. These users are, therefore, likely to stabilize on a throughput lock-in, reinforced by the relative accuracy and fluency of LLM outputs across a wide spectrum of domains. Figure ~\ref{fig:path_b} graphically depicts the dynamics of throughput lock-in.

\begin{figure}[t!]
\centering
\includegraphics[width=\linewidth]{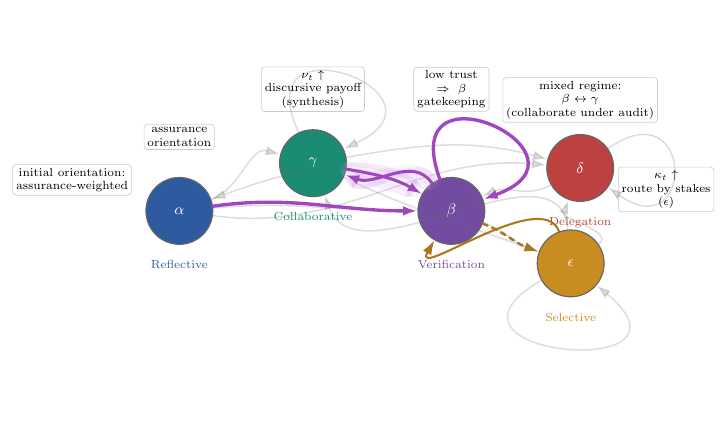}
\caption{Canonical Path C (mixed-assurance): low-trust users adopt \(\beta\) and continue using the system when \(\beta \leftrightarrow \gamma\) yields discursive payoff; burden can induce stake-based routing (\(\epsilon\)).}
\label{fig:path_c}
\end{figure}

\paragraph{Mixed Assurance}
The mixed assurance equilibrium arises when users at the outset exhibit an assurance orientation toward AI outputs characterized by low-trust and caution. These users are not likely to accept LLM outputs at face value but subject them to verification chains against trusted sources, triangulation and careful examination. Across repeated interactions, these users gravitate toward a collaborative dynamic marked by selective verification of high-stakes outputs and continual modification of outputs. These users are likely to settle on a stable regime marked by selective auditing and verification checks on LLM outputs, whether these are final products or inputs to human-generated task. Figure ~\ref{fig:path_c} graphically depicts the dynamics of verification gating.  

Now that we have worked out the individual decision-theoretic space, we turn to several salient game-theoretic scenarios we believe beset all human-AI interaction and collaboration: (a) principal-agent games, (b) signaling games, and (c) Bayesian persuasion games. 

\subsection{Strategic Communication Games}
Following extended mind theory and cognitive coupling theory \cite{clark_chalmers_1998, menary_2010, clark2025extending} within enactivist cognition \cite{Thompson2010}, we conceptualize human-AI interaction as constituting a communicative coupling that emerges through repeated and ongoing interactions. Human-AI communicative coupling is a distinct and novel dimension of cognitive coupling that straddles the space of cognitive extensions as prosthetics of individual action and communicative action as a domain of social interaction that involves the exchange of validity claims and the justification of norms between communicative agents.  As an extension of human-computer cognitive coupling \cite{hila2026enactivist}, human-AI coupling entails a distribution of cognitive resources within the coupled system. This conceptualization has strong precedent in HCI, where the human agent offloads cognitive processing power into the technology to which they are coupled. This is evident with calculators, for example, but paradigmatically in human-computer coupling where the computer functions as an extension of human agent memory storage and information processing power. Therefore, one side of the equation frames human-AI coupling as continuous with cognitive coupling theory in HCI. The other side of the equation, however, namely human-AI communicative coupling, extends beyond the prosthetics of human cognition framing and forays into symmetrical communicative action, where two agents with relatively equal linguistic competence exchange both information and reasons for belief and action. 

We argue that the second framing engenders game-theoretic-adjacent interactions between the human and AI counterparts that roughly follow the communicative dynamics explored and formalized in information economics as a communicative extension of game theory. Information economics begins with Harsanyi's \textit{Games with Incomplete Information Played by Bayesian Players} \cite{harsanyi1967games1}, where he extends game theory from deterministic scenarios where payoffs are known by players to games where players possess incomplete information. Harsanyi's chief contribution consisted in solving the infinite regress of belief expectations between agents by assigning agents a common probabilistic prior about each other's hidden states such as preferences or type \cite{harsanyi1967games1}. Harsanyi's formalism introduced (a) chance as a variable that probabilistically assigns agent hidden states and (b) a common prior that converted incomplete information to probabilistic or imperfect information \cite{harsanyi1967games1}. Auman extended Harsanyi's Bayesian solution to iterated Bayesian games where players ``leak information" through repeated interactions that factor into Bayesian updates \cite{aumann1995repeated}. However, neither of these models incorporated explicit information transmission into the dynamics of interaction until Spence introduced signaling as a type of manipulable observable information that enables senders to maximize their utility \cite{spence1973job}. Explicit communication in the form of costless and costly strategic information transmission marked an extension of communicative game theory by introducing information transmission to incomplete information games between agents \cite{crawford1982strategic, spence1973job, farrell1996cheap, cho1987signaling}. Following reverse-adaptation theory \cite{winner1978autonomous, clark_2008} and communicative game theory \cite{harsanyi1967games1, aumann1995repeated}, we explore human-AI communicative action that spans the following spectrum of communication games: principal-agent games, signaling games, and Bayesian persuasion games. 

Across these communication games, we identify four overarching variables that govern and regulate the evolving communicative dynamics. Both human and AI agents possess incomplete information, with the relation being marked by an information-asymmetry where the informational distribution skews positively to AI agents. It is important to qualify that this information asymmetry does not entail a one-to-one correspondence to a knowledge asymmetry, given that conditions for knowledge differ between internalist and reliabilist accounts \cite{Hila2025}, with AI systems more plausibly satisfying the latter criteria. These two conditions shape and constrain the communicative exchange between human and AI agents. In addition, each agent possesses a variable axis of knowledge about the other, namely AI alignment for the AI agent and AI literacy for the human agent. The alignment of the LLM chatbot shapes the types of signaling it transmits to the human agent, with some chatbots displaying extreme confidence in their outputs and others exhibiting degrees of diffidence and uncertainty \cite{ji2023ai}. Further, evidence shows that human agent AI literacy influences their ontological designation of the AI agent, their degree of trust, and the types and effectiveness of use \cite{huang2024influence}. Therefore, we submit incomplete information, information asymmetry, AI alignment and AI literacy as four dimensions that constrain the communicative games we explore below. 

\paragraph{Principal-Agent Games} The principal-agent problem refers to conflicts of interest arising in arrangements in which one party, the principal, delegates an action to another party, the agent, who executes the action on the principal's behalf \cite{ross1973agency, jensen1976theory, mirrlees1976optimal, holmstrom1979moral}. Human systems abound with principal-agent relationships such as citizens and representatives, shareholders and the board of directors, and employer and employee, to name a few. We argue that prototypical human-AI communication follows a principal-agent relationship, where the human agent assumes the role of the principal that delegates a task to the AI, who in turn assumes the role of an agent acting on the human's behalf. When prompting the AI chatbot, the human agent's intention is to obtain an output or outcome that matches their explicit or latent objectives. A common source of friction in human principal-agent arrangements concern incentives for the agent to defect from the principal's objectives. In the case of human-AI interaction, AI defection must be conceptualized less as an intention of the system and more of a system disposition in context of the LLM's hyperparameters, data provenance, fine-tuning, pre-alignment, and alignment phases. 

Conflicts of interest between human and AI arise when there's misalignment between human intentions and AI outputs. Even though AI alignment techniques such as RLHF aim to shape AI outputs that are helpful and harmless, the AI system can always misinterpret or misunderstand the intention of the human agent from the input prompt. Highly aligned, high context-window LLM chatbots and highly AI literate human agents trained in effective prompting strategies mitigate risks of human-AI principal-agent problems. On the other hand, poorly aligned, low context-window LLMs in conjunction with low AI literacy human agents exacerbate human-AI principal-agent problems. We now turn to how these problems arise in specific communicative scenarios.  

\paragraph{Signaling Games} We define two types of signaling games in the context of the principal-agent relationship we introduced: (a) cheap talk and (b) costly signaling. Cheap talk refers to messages that a sender conveys to a receiver \cite{crawford1982strategic, farrell1996cheap}, whereas costly signals refer to high-quality output cues \cite{spence1973job}. Because human-AI interaction games entail incomplete information from both parties, signaling can occur in two ways: (a) high-quality outputs and (b) facsimiles of high-quality outputs that are, in fact, defective. Unlike in human scenarios, all costly signaling by an AI system, such as programmable scripts, can be diluted by cheap talk such as the semblance of competence or hallucinations. Because AI outputs can signal competence and trustworthiness without actually being accurate, they shift the verification burden onto the human agent, namely the principal. The cost of verification for the human agent is proportional to (a) their degree of competence in the knowledge domain at hand and (b) their AI literacy. 

Given these distinctions within human-AI communication, we posit three types of communicative patterns representing a trustworthiness scale from lowest to highest: (a) cheap talk, (b) verifiable disclosure, and (c) Bayesian persuasion. Unstructured human-AI communication, where the human receiver imposes little to no verifiability and sourcing standards, is laden with inherent structural risks of cheap talk that follow from emergent behaviors such as sycophancy and reward hacking \cite{eisenstein2023helping, sharma2023towards, chen2025selfaugmented}. Consequently, high stakes interactions and usage cannot rely on subjective determinations of model trustworthiness. To mitigate these risks, human users and organizations must impose verifiable disclosure constraints and approximate Bayesian commitments on AI agents. These constrains can be broadly divided into pre and post-usage. The burden of pre-usage constrains fall on AI model designers and include strategies in supervised fine-tuning and alignment stages such as (a) rewarding epistemic sincerity, (b) constraining output formats to evidence-bearing schemas, and (c) incorporating tools such as RAG and data provenance policies. These variables, however, lie outside users' control. As such, we focus on two types of post-usage strategies as communicative constraints that steer models toward high-quality outputs in high-cost scenarios: verifiable disclosure and Bayesian commitments.  

\paragraph{Verifiable Disclosure Games} Verifiable disclosure refers to a subdomain of information economics developed by Grossman and Milgrom \cite{grossman1981informational, milgrom1981good}, who argued that firms have an overriding incentive toward product-quality disclosure because the alternative implies a ``worst-case" inference by rational consumers, triggering an unraveling through verifiable disclosure signals. In the context of human-AI interaction, the goods exchanged constitute information and compute. In our case, we define verifiable disclosure as AI-produced information artifacts that minimize the cost of a user's independent verification burden. Incentive alignment between users and AI designers materializes under conditions where users seek reliable outputs whereas AI companies in a competitive market seek to maximize subscriptions and premium usage. However, in an environment saturated with substitutable goods, misalignment ensues when AI designers can maximize usage through trustworthiness cues and sycophancy rather than verifiable disclosure. In general, as the proportion of users increases, AI firms have an incentive to maximize trustworthiness cues rather than trustworthiness as such through verifiable disclosure. This misalignment potential follows from the reverse adaptation theory we have propounded as a structural problem besetting human-AI collaboration. AI system designers have an overriding incentive to maximize usage as a proxy for profit, not user utility maximization. Human agents can deploy several strategies to elicit verifiable outputs from AI agents: (a) evidence-gated prompting, (b) adversarial and agentic verification loops, and (c) provenance traces. 

\paragraph{Bayesian Persuasion Games}
Bayesian Persuasion Games constitute an extension of information economics where interacting agents transmit information structures to influence the beliefs and actions of other rational agents under conditions of asymmetric information \cite{kamenica2011bayesian}. The communicative exchange between human and LLM chatbots are particularly liable to Bayesian persuasion games. However, in contrast to human communication, in human-AI interaction both the signal and the product come packaged into the same output, making the communicative exchange especially liable to distortion. The signal regime can be analyzed into a nested dependency of quasi-commitments. Designers commit ex-ante to an engagement maximization strategy shaped by RLHF where trustworthiness cues override epistemic objectivity. The model's architecture, furthermore, entails hallucinations as a structural possibility, which can be rendered invisible through AI trustworthiness cues such as signal fluency and confidence. Furthermore, users can leverage the model's complex probability density through prompt engineering to assign the model personas or information-producing patterns with context-dependent commitments and obligations. 

To map how Bayesian persuasion fits into human-AI interaction we must consider the human-AI information asymmetry, mutual uncertainty, AI alignment and AI literacy as variables that govern the assignment of priors and structure emerging posteriors across both agents. In empirical scenarios, human agents likely begin with non-neutral assumptions, known as informative priors, about AI trustworthiness and capabilities. Consequently, we can map out Bayesian belief updates as stochastic belief-trajectories that commence either with either informed or neutral priors. When mapping belief trajectories with informed priors, the source of the prior and the users' doxastic attitudes matter. Users with high AI literacy may have structural understanding and be doxastically rigid, whereas users with low AI literacy may harbor culturally dominant beliefs and be doxastically flexible, and vise versa. Both types of users may begin with variable trustworthiness assignments, but high AI literacy users are likely to have more grounded beliefs than low literacy users. Nevertheless, both types of users are liable to Bayesian persuasion whereby across repeated interactions, they experience trustworthiness increase as a consequence of the structural conflation of cheap-talk and verifiable disclosure. At this juncture, reverse adaptation theory can inform the interaction. Because AI designers have an incentive to maximize engagement, the conjunction of model confidence and its information asymmetry with human users can function to shape users' posterior beliefs, including unknowingly injecting false beliefs. While all user types are liable to these effects, low literacy users are most at risk \cite{huang2024influence}. Because designers can commit their fine-tuning and alignment stages to maximize trustworthiness cues rather than epistemic detachment, the model can systematically exploit users' Bayesian update mechanism. In conjunction with fluency, user belief affirmation known as sycophancy can structure Bayesian updates toward confirmation bias rather than grounded beliefs.

\paragraph{Coupled Human-AI $\longleftrightarrow$ Human-AI Interactions} If thus far we have considered signaling and messaging between humans and AI within a principal-agent dynamic, in this section we transition to signaling and messaging between human-AI coupled agents. Messaging and signaling in a world where human agents know their own human-AI ``type" or strategy but do not know others' strategy constitutes a game with incomplete or imperfect information if we assign common priors to agents that update on repeated interactions.

\subsection{Collective Equilibria}
We now proceed to scaffold our human-AI coupled individual strategies to collective equilibria. In mapping possible collective equilibria, we consider three plausible, theoretical scenarios that converge at the cleavage of local-to-global dynamics. The first scenario considers how the strategic actions and evolutionary trajectories of individual human agents scale into collective outcomes. The second scenario considers how human-AI coupled agents interact with each other through signaling, information leakage, and communicative action. The third scenario considers how authoritative norms can be conducive to shifting a collectively suboptimal equilibrium toward an optimal one. 

The second model introduces contingencies in individual human-AI interaction strategies that absorb influential signals from other users. Information leakage occurs when across iterative Bayesian interactions players reveal information about their type. In the case of human-AI interaction, this can be information artifacts, outcomes, and action patterns that indirectly reveal the user's strategy. On the other hand, strategic disclosure and nondisclosure through intentional signaling can take the form of ``cheap talk" such as asserting ``I don't use AI" to save face, or calculated misdirection such as partial disclosure by asserting that they use it only as their assistant. In each scenario, agents are expected to take hidden actions that implement their utility-maximization strategy. Signals can propagate through local networks, influencing local norms or strategy-signaling behaviors. For example, if some agents in a network assert that they don't use AI, this can signal to other agents in the network one of two ambiguous meanings: (a) they choose not to disclose their usage or (b) that usage, at least varying degrees of it, may be frowned upon in the network. Depending on the interpretation, users may be influenced to (a) conceal disclosure or (b) become deterred to use AI in the relevant group-context. Against ambiguous signals, agents are likely to pursue their individually optimal strategies through hidden actions. However, following Habermas \cite{habermas2007reason}, it is important to consider that speech signals carry norm-binding or norm-negotiation force. For example, if the group carries a sustained discussion on appropriate usage, the reasoned discursive claims can impose normative commitments on the group. This is because, according to Habermas, interpersonal speech implies a mutual intention of valid consensus generation \cite{habermas2007reason}.

In line with the foregoing, we advance that scaled human-AI interaction poses an array of collective action problems as well as coordination opportunities. Collective action problems paradigmatically refer to situations in which individuals stand to gain from cooperation but fail to do so \cite{olson1965logic, hardin1982collective}. We highlight two types of problems: prisoner's dilemmas and coordination problems. Prisoner's dilemmas refer to situations that embody conflicts of interest between individual and group interests \cite{axelrod2006evolution}. Coordination problems, on the other hand, refer to situations where group members have the same objective but fail to coordinate their plans to achieve it \cite{heath2009communicative}. The central prisoner's dilemma centers on the conflict between human agents maximizing delegative human-AI interaction strategies at the expense of responsible, verificationist strategies, depicted in table ~\ref{tab:prisoners}. When scaled to the n-payer prisoner's dilemma, the collective optimum of responsible strategies yields collective outcomes of robust epistemic chains and trustworthy products. However, mutual cooperation at the societal scale becomes liable to Olson's free rider problem where cooperation payoffs are marginally miniscule in the short term, even though cooperation benefits society in the long term. 

Like the prisoner's dilemma scenarios, coordination challenges can be framed at the interactive group and collective levels. Unlike the prisoner's dilemma, the central coordination challenge lies in adopting universal norms that produce optimal collective equilibria. In particular, we frame coordination as the collective objective of high-transparency and high provenance standards toward optimal group outcomes and no disclosure and low provenance as exacerbating coordination, illustrated in table ~\ref{tab:coordination}. In small groups, such as a school project or a small start up, everyone has a common objective of producing a high-quality product. Coordination in such groups can be achieved through explicit communication. In particular, following communicative action theory, we argue that explicit communication imposes normative obligations on members \cite{habermas2007reason, heath2009communicative}. Even though we have distinguished conflict of interest and coordination scenarios in theory, real-life scenarios embody complex mixtures of the two. In small groups, small communication costs and large defection costs make coordination and cooperation more probable. However, as the scale of collectivity increases, both cooperation and coordination become difficult to achieve without additional crutches \cite{olson1965logic, ostrom1990governing}. 

Consequently, we propose three aggregation principles that shape evolutionary trajectories at the collective scale: (a) the aggregation of microbehaviors without signaling, (b) microbehaviors with signaling and communication and (c) institutional effects. We posit that emergent global and local equilibria from scaled human-AI interaction are likely to embody complex mixtures of these three aggregation principles. Aggregate effects of individual strategies without signaling or communication are likely to exacerbate cooperation and coordination problems. Information leakage and local communication networks are likely to seed local norms that modify individual, decision-theoretic strategies. And lastly, institutional interventions can function to iron out inconsistencies between local norms and individual pay-off maximizing strategies through top-down mechanisms of norm construction and diffusion. We proceed to formalize each of these aggregation principles through a probability model for network population dynamics. 


\begin{table}[t]
\centering
\small
\renewcommand{\arraystretch}{1.25}
\resizebox{\textwidth}{!}{%
\begin{tabular}{lcc}
\toprule
 & \textbf{Player 2: C (Responsible)} & \textbf{Player 2: D (Delegative)} \\
\midrule
\textbf{Player 1: C (Responsible)} & $(R,R)$ & $(S,T)$ \\
\textbf{Player 1: D (Delegative)}  & $(T,S)$ & $(P,P)$ \\
\bottomrule
\end{tabular}
}
\caption{Prisoner’s Dilemma for responsible vs delegative AI use. $C$ denotes selective offloading/verification when required; $D$ denotes throughput-maximizing delegation with functional satisficing outputs. Payoffs satisfy $T>R>P>S$.}
\label{tab:prisoners}
\end{table}

\begin{table}[t]
\centering
\small
\renewcommand{\arraystretch}{1.25}
\resizebox{\textwidth}{!}{%
\begin{tabular}{lcc}
\toprule
 & \textbf{Player 2: H (Disclose + Provenance)} & \textbf{Player 2: L (No disclosure / low provenance)} \\
\midrule
\textbf{Player 1: H (Disclose + Provenance)} & $(A,A)$ & $(C,D)$ \\
\textbf{Player 1: L (No disclosure / low provenance)} & $(D,C)$ & $(B,B)$ \\
\bottomrule
\end{tabular}
}
\caption{Coordination game for small-group AI-use norms. $H$ is a high-integrity norm package (disclosure + provenance requirements); $L$ is a low-integrity package (minimal disclosure/provenance, speed-first). Payoffs satisfy $A>B$ and $A>D$, $B>C$ (matching beats mismatching).}
\label{tab:coordination}
\end{table}

\paragraph{Aggregation Principle I: Distributional Scaling of Microbehaviors.}
Let
\[
R=\{r_1,r_2,r_3\}
\]
denote the three canonical individual end-states, defined as metastable user regimes, and let
\[
\pi_t \in \Delta(R), 
\qquad
\pi_t(r_k)=\Pr(r_t=r_k),
\qquad
\sum_{k=1}^3 \pi_t(r_k)=1.
\]
$R$ defines the sample space, $\pi_t$ defines the distribution of probabilities in time normalized to the unit interval, and $\Omega_t$ defines the macroscopic state variables. This first aggregation principle treats collective outcomes as a function of the population distribution over individual regimes:
\[
\Omega_t = F(\pi_t).
\]
Here, macrobehavior is generated by scaled microbehavior under a given type distribution, without yet incorporating communication or institutional norm diffusion.

\paragraph{Aggregation Principle II: Norm Diffusion through Signaling and Communication.}
Let agents be embedded in a local interaction network \(G=(V,E)\). Each agent \(i\) occupies an individual strategy \(s_i(t)\in S\), but strategy transitions are influenced by neighboring signals such as disclosures, concealments, and observed outputs:
\[
s_i(t+1)=\Phi\!\Big(s_i(t),\,z_i(t),\,\sum_{j\in N(i)} w_{ij}\,\psi(s_j(t),d_j(t),y_j(t))\Big).
\]
In the above equation, \(s_i(t+1)\) denotes the future state of agent $i$, \(s_i(t)\) the agent's current state, \(z_i(t)\) the agent's individual preferences, \(s_j(t)\) direct neighbors in the agent's network, \(d_j(t)\) agent network disclosure behavior and \(y_j(t)\) network observable products or outputs. The transition dynamics from the current strategy to a future strategy are a function of the social influence observed from the agent's immediate social network denoted by the weighted sum of the neighbors' ($j$) behaviors and signals: \(\sum w_{ij}\ \psi\), where $\psi$ denotes the network signal, and $\phi$ the transition function. Under this aggregation principle, local signaling and communicative exposure reshape individual strategies, producing endogenous local norms, including norms of disclosure, concealment, verification, or routine delegation.

\paragraph{Aggregation Principle III: Institutional Constraint and Regime Selection.}
Let \(I_t\) denote the institutional environment laden with disclosure rules, provenance requirements, audits, penalties, and workflow mandates. Institutional structure acts on both the global distribution of individual regimes and the locally emergent communicative equilibria:
\[
\Omega_t = H(\pi_t,\;G,\;I_t).
\]
In this formulation, $\Omega_t$ represents the aggregate system state, $H$ is the aggregation function, $\pi_t$ signifies the distribution of individual agent strategies, $G$ denotes the network topology of interactions, and $I_t$ provides the exogenous institutional constraints. Institutional effects do not eliminate utility-maximizing behavior, but they alter the payoff structure under which local norms and global distributions interact. In this sense, institutions function as regime-selection devices: they can dampen suboptimal local equilibria, reinforce responsible norms, and partially align local communicative dynamics with global epistemic standards.

\paragraph{Equilibrium: Delegation Lock-In} Following Arthur's lock-in theory \cite{arthur1989competing}, we argue that scaling individual strategies without communicative interventions can lead to suboptimal lock-in. Contrary to Arthur, we argue that lock-in doesn't necessarily lead to increasing returns. To make this claim plausible, let's consider how lock-in through path dependencies can impose onerous switching costs that can drag responsible human agents into a coercive vortex of low-quality delegation. Consider existing suboptimal technological lock-ins whose nested dependencies make switching extremely costly. A user has personal privacy-related incentive to switch from their current integrated utility platform, which integrates cloud storage, compute environments and application services, into a cheaper, privacy-robust yet disaggregated suite of services. However, system familiarity, interoperability, and low collaboration enlisting costs make switching costly and contribute to lock-in. At the sociological scale, a major factor contributing to collective lock-in are critical mass thresholds. Delegation lock-in refers to an equilibrium where responsible users prone to strategies $\alpha$, $\beta$, and $\epsilon$ are pulled into $\delta$ -- the instrumental delegation strategy -- because not adopting the high-throughput option puts them at a disadvantage when the majority have opted-in. Delegation lock-in can be a temporarily stable strategy but can become unstable if lack of oversight leads to a pooling equilibrium where products are indistinguishable and undetected errors proliferate. 

\paragraph{Equilibrium: Stratified Assurance} A plausible equilibrium not dominated by instrumental delegation strategies includes one in which a stable distribution of mixed user strategies or user types emerge. This equilibrium recognizes that, besides being in flux, individual strategies are constrained by user types and subject to signaling and communication effects that can induce users to accept obligations and commitments. We classify users into throughput oriented users who prioritize efficiency (dominated by the $\delta$ strategy), assurance-oriented users who prioritize verification (dominated by the $\beta$ strategy), and reflective-reasoning users who prioritize understanding (dominated by $\alpha$). Consequently, we define stratified assurance as a mixed, population-level equilibrium characterized by the interaction of heterogeneous user types and local communicative environments, such that a persistent minority of reflective ($\alpha$) and verification-oriented ($\beta$) users stave-off full convergence into a delegation lock-in equilibrium. Reflective users are motivated to use AI to increase their understanding and improve their epistemic states. Verificationist users, whether reflectively motivated or not, are unlikely to accept AI outputs at face-value alone. It is important to qualify here that the second aggregation principle can lead to either delegation lock-in or stratified assurance. Lock-in equilibria require signaling such as the observation that others in one's network are exploiting usage, thereby incentivizing adoption from observed abstention costs. However, we posit that local communicative action is likely to produce complex effects, where users take cues from their peers about norms of usage such that subjective utility maximization becomes tempered by emergent social norms and prospects of negative sanctioning. As such, signaling and communication in local networks can influence individual strategies by either (a) amplifying delegation or (b) amplifying responsibility. Our contention is that (b) follows from communicative action, whereas (a) follows from indirect signaling. Because user types are unevenly distributed, yet subject to varying degrees of recalcitrance, and local networks subject to varying degrees of communicative standards, mixed equilibria with emergent pockets of responsible use form a plausible evolutionary outcome.

\paragraph{Equilibrium: High-Integrity Provenance} The third plausible equilibrium takes into account the top-down effects of institutional and organizational adoption. As the adoption of Generative AI reaches critical mass, large-scale effects on productivity and product quality will become apparent. As the effects of the distribution of initial strategies become observable on the collective scale, organizations and institutions are likely to develop and impose evidence-based usage models, rules, and norms. The emergence of institutional and organizational norms are, at least at first, likely to be heterogeneous, but may converge on large-scale consensus on certain usage parameters. The generation and institution of norms constitute a top-down mechanism that regulates both adoption and usage patterns. If the first two equilibria result from the first and second aggregation principles, the high-integrity provenance equilibrium follows from the third aggregation principle. The dissemination of institutional norms can function as a mechanism that harmonizes emerging local norms with distributed global usage patterns. In other words, institutional controls can reduce free-rider problems stemming from n-agent, multi-shot prisoner's dilemmas by constraining or fettering individual utility maximization to collectively desirable standards. While they are unlikely to eliminate free-ridership nor homogenize the strategic landscape, they restructure payoffs through organizational rewards and punishments. 

\subsection{Sociotechnical Lock-In}
Given the foregoing considerations, we refine Arthur's linear equation for technological lock-in by making payoffs proportional to the cost of the process in obtaining desired payoffs. At first, AI adoption appears to maximize individual utility by minimizing the effort expended for the same outcomes. In ideal conditions, users forego epistemic effort while conserving or improving the quality of the outcomes. However, in concrete scenarios, maximum delegation can lead to the propagation of undetected errors in $P_A$, the success rate of path A. Simultaneously, delegation can lead to cognitive atrophy and the erosion of the ability of human epistemic agents from evaluating outcomes. Therefore, while the maximum delegation strategy at first maximizes individual utility, long-term it scales to collective outcomes that are undesirable. 

\[
U_A(n_A)=\frac{P_A(n_A)}{C_A}+r\,n_A
\]

\[
\frac{\partial P_A}{\partial n_A}<0
\]

In this formulation, \(U_A(n_A)\) denotes the individual utility of adopting path \(A\) at adoption level \(n_A\). The term \(P_A(n_A)\) denotes the realized outcome quality under path \(A\), while \(C_A\) denotes the process cost associated with that path, so that the ratio \(P_A(n_A)/C_A\) captures the outcome achieved relative to the effort expended. The term \(r\,n_A\) represents the Arthurian coordination return from adopting the same path as others, where $n$ denotes the number of users and $r$ the coordination gain coefficient. Unlike the canonical Arthurian formulation, in which returns increase linearly with adoption through network effects, our model allows the intrinsic quality of outputs \(P_A(n_A)\) to vary with adoption in proportion to the elided process cost. Therefore, while \(P_A(n_A)\) may increase at low levels of adoption, it begins to deteriorate beyond a critical threshold \(n_A^\dagger\), namely when scaled use proliferates errors while simultaneously diminishing evaluative capacity in human agents. Thus, AI adoption can initially increase individual utility through low process cost and coordination gains. However, in the long run, mass adoption produces prisoner’s dilemma collective action problems where rational individual adoption persists while collective outcomes decline.

\section{Conclusion}
Our paper makes several important contributions. We have shown that human-AI interaction deeply complexifies the landscape of strategic action by introducing new collective action problems. To this effect, we have shown that evolutionary game theory is consistent with multiple metastable equilibria of collective human-AI strategies. We have argued that suboptimal equilibria follow as a consequence of unbridled individual utility maximization strategies that human agents are liable to adopt. We have further shown that intrinsic value-differentials between agents are likely to yield mixed-strategy equilibria dominated by delegation yet alloyed with reflective and verificationist strategies. Further, we have also shown that pure strategies are likely to be unstable across recurring interactions as users learn to optimize the human-AI delegation-verification dilemma. However, because individual optima vary across user types, mixed-strategies are unlikely to mitigate the risks of the dissemination of errors, the erosion of AI-generated payoffs, and the atrophy human evaluation capability. To this effect, we have extended Arthur's theory of lock-in by arguing that critical-mass network effects from AI adoption are not tethered to a monotonous function of increasing returns but rather something we have termed \textit{sociotechnical lock-in}, a collective state in which high exit costs anchor users to a self-reinforcing normative arrangement, despite the aggregate of individual decisions increasing yielding diminishing returns. Consequently, we have presented a plausible scenario where initial mass adoption produces inflexible lock-in in spite of individual payoffs atrophying. We have provided two solutions: (a) a bottom up communicative solution where local network communication seeds suboptimality offsetting norms and (b) a top-down solution where organizations and institutions impose norm-bound deontic commitments on users toward adopting responsible strategies that offset collective suboptimality both at the organizational and, a fortiori, societal scales.

\printbibliography

@article{clark2025extending,
  author    = {Andy Clark},
  title     = {Extending minds with generative {AI}},
  journal   = {Nature Communications},
  volume    = {16},
  pages     = {4627},
  year      = {2025},
  doi       = {10.1038/s41467-025-59906-9},
  url       = {https://www.nature.com/articles/s41467-025-59906-9}
}

@book{simon1996,
  author    = {Simon, Herbert A.},
  year      = {1996},
  title     = {The Sciences of the Artificial},
  edition   = {3},
  publisher = {MIT Press},
  address   = {Cambridge, MA}
}

@book{heath2011,
  author    = {Heath, Joseph},
  year      = {2011},
  title     = {Following the Rules},
  publisher = {Oxford University Press},
  address   = {New York, NY}
}

@book{bostrom2016,
  author    = {Bostrom, Nick},
  year      = {2016},
  title     = {Superintelligence: Paths, Dangers, Strategies},
  publisher = {Oxford University Press},
  address   = {Oxford, England}
}

@article{clark_chalmers_1998,
  author  = {Clark, Andy and Chalmers, David J.},
  title   = {The Extended Mind},
  journal = {Analysis},
  year    = {1998},
  volume  = {58},
  number  = {1},
  pages   = {7--19},
  doi     = {10.1093/analys/58.1.7}
}

@book{clark_2008,
  author    = {Clark, Andy},
  title     = {Supersizing the Mind: Embodiment, Action, and Cognitive Extension},
  publisher = {Oxford University Press},
  address   = {Oxford, UK},
  year      = {2008}
}

@article{clark_2013_whatever_next,
  author  = {Clark, Andy},
  title   = {Whatever Next? Predictive Brains, Situated Agents, and the Future of Cognitive Science},
  journal = {Behavioral and Brain Sciences},
  year    = {2013},
  volume  = {36},
  number  = {3},
  pages   = {181--204},
  doi     = {10.1017/S0140525X12000477}
}

@book{menary_2010,
  editor    = {Menary, Richard},
  title     = {The Extended Mind},
  publisher = {MIT Press},
  address   = {Cambridge, MA},
  year      = {2010}
}

@article{jarrahi_lutz_newlands_2022,
  author  = {Jarrahi, Mohammad Hossein and Lutz, Christoph and Newlands, Gemma},
  title   = {Artificial intelligence, human intelligence and hybrid intelligence based on mutual augmentation},
  journal = {Big Data \& Society},
  year    = {2022},
  volume  = {9},
  number  = {2},
  pages   = {1--6},
  doi     = {10.1177/20539517221142824},
  url     = {https://journals.sagepub.com/doi/full/10.1177/20539517221142824}
}

@inproceedings{wang_churchill_maes_fan_shneiderman_shi_wang_2020,
  author    = {Wang, Dakuo and Churchill, Elizabeth F. and Maes, Pattie and Fan, Xiangmin and Shneiderman, Ben and Shi, Yuanchun and Wang, Qianying},
  title     = {From Human–Human Collaboration to Human–AI Collaboration: Designing AI Systems That Can Work Together with People},
  booktitle = {Extended Abstracts of the 2020 CHI Conference on Human Factors in Computing Systems},
  year      = {2020},
  pages     = {1--6},
  doi       = {10.1145/3334480.3381069},
  url       = {https://dl.acm.org/doi/pdf/10.1145/3334480.3381069}
}

@article{arthur1989competing,
  author  = {Arthur, W. Brian},
  title   = {Competing Technologies, Increasing Returns, and Lock-In by Historical Events},
  journal = {The Economic Journal},
  year    = {1989},
  volume  = {99},
  number  = {394},
  pages   = {116--131},
  doi     = {10.2307/2234208}
}

@book{schelling1978micromotives,
  author    = {Schelling, Thomas C.},
  title     = {Micromotives and Macrobehavior},
  publisher = {W. W. Norton},
  address   = {New York, NY, USA},
  year      = {1978}
}

@book{olson1965logic,
  author    = {Olson, Mancur},
  title     = {The Logic of Collective Action: Public Goods and the Theory of Groups},
  publisher = {Harvard University Press},
  address   = {Cambridge, MA, USA},
  year      = {1965}
}

@book{winner1978autonomous,
  author    = {Langdon Winner},
  title     = {Autonomous Technology: Technics-out-of-Control as a Theme in Political Thought},
  year      = {1978},
  publisher = {MIT Press},
  note      = {Original work published 1977}
}

@book{clark2003naturalborncyborgs,
  author    = {Andy Clark},
  title     = {Natural-Born Cyborgs: Minds, Technologies, and the Future of Human Intelligence},
  year      = {2003},
  publisher = {Oxford University Press},
  address   = {Oxford / New York},
  isbn      = {9780195148664}
}

@book{axelrod2006evolution,
  author    = {Axelrod, Robert},
  title     = {The Evolution of Cooperation},
  edition   = {Revised},
  publisher = {Basic Books},
  address   = {New York, NY, USA},
  year      = {2006},
  isbn      = {9780465005644},
  url       = {https://www.amazon.com/Evolution-Cooperation-Revised-Robert-Axelrod/dp/0465005640}
}

@book{ostrom1990governing,
  author    = {Ostrom, Elinor},
  title     = {Governing the Commons: The Evolution of Institutions for Collective Action},
  publisher = {Cambridge University Press},
  address   = {Cambridge, UK},
  year      = {1990},
  isbn      = {9780521405998},
  url       = {https://www.amazon.com/Governing-Commons-Evolution-Institutions-Collective/dp/0521405998}
}

@article{granovetter1978threshold,
  author  = {Granovetter, Mark S.},
  title   = {Threshold Models of Collective Behavior},
  journal = {American Journal of Sociology},
  year    = {1978},
  volume  = {83},
  number  = {6},
  pages   = {1420--1443},
  doi     = {10.1086/226707},
  url     = {https://doi.org/10.1086/226707}
}

@book{heath2009communicative,
  author    = {Heath, Joseph},
  title     = {Communicative Action and Rational Choice},
  publisher = {MIT Press},
  address   = {Cambridge, MA, USA},
  year      = {2009},
  isbn      = {9780262582247},
  url       = {https://mitpress.mit.edu/9780262582247/communicative-action-and-rational-choice/}
}

@book{heath2014enlightenment,
  author    = {Heath, Joseph},
  title     = {Enlightenment 2.0: Restoring Sanity to Our Politics, Our Economy, and Our Lives},
  publisher = {HarperCollins},
  address   = {Toronto, ON, Canada},
  year      = {2014},
  isbn      = {9781443422536},
  url       = {https://www.amazon.com/Enlightenment-2-0-Joseph-Heath/dp/1443422533}
}

@book{Thompson2010,
  author    = {Thompson, E.},
  title     = {Mind in Life},
  publisher = {Belknap Press},
  year      = {2010},
  address   = {London, England},
}

@incollection{hila2026enactivist,
  author    = {Hila, Angjelin},
  title     = {An Enactivist Approach to Human-Computer Interaction: Bridging the Gap Between Human Agency and Affordances},
  booktitle = {HCI International 2025 – Late Breaking Papers: 27th International Conference on Human-Computer Interaction, HCII 2025, Gothenburg, Sweden, June 22–27, 2025, Proceedings, Part I},
  editor    = {Kurosu, Masaaki and Hashizume, Ayako},
  series    = {Lecture Notes in Computer Science},
  volume    = {16331},
  publisher = {Springer},
  address   = {Cham, Switzerland},
  year      = {2026},
  pages     = {28--48},
  doi       = {10.1007/978-3-032-12657-3_3}
}

@article{crawford1982strategic,
  author  = {Crawford, Vincent P. and Sobel, Joel},
  title   = {Strategic Information Transmission},
  journal = {Econometrica},
  year    = {1982},
  volume  = {50},
  number  = {6},
  pages   = {1431--1451},
  doi     = {10.2307/1913390}
}

@article{farrell1996cheap,
  author  = {Farrell, Joseph and Rabin, Matthew},
  title   = {Cheap Talk},
  journal = {Journal of Economic Perspectives},
  year    = {1996},
  volume  = {10},
  number  = {3},
  pages   = {103--118},
  doi     = {10.1257/jep.10.3.103}
}

@article{spence1973job,
  author  = {Spence, Michael},
  title   = {Job Market Signaling},
  journal = {Quarterly Journal of Economics},
  year    = {1973},
  volume  = {87},
  number  = {3},
  pages   = {355--374},
  doi     = {10.2307/1882010}
}

@article{cho1987signaling,
  author  = {Cho, In-Koo and Kreps, David M.},
  title   = {Signaling Games and Stable Equilibria},
  journal = {Quarterly Journal of Economics},
  year    = {1987},
  volume  = {102},
  number  = {2},
  pages   = {179--221},
  doi     = {10.2307/1885060}
}

@article{kamenica2011bayesian,
  author  = {Kamenica, Emir and Gentzkow, Matthew},
  title   = {Bayesian Persuasion},
  journal = {American Economic Review},
  year    = {2011},
  volume  = {101},
  number  = {6},
  pages   = {2590--2615},
  doi     = {10.1257/aer.101.6.2590}
}

@article{harsanyi1967games1,
  author  = {Harsanyi, John C.},
  title   = {Games with Incomplete Information Played by {“Bayesian”} Players, Part I: The Basic Model},
  journal = {Management Science},
  year    = {1967},
  volume  = {14},
  number  = {3},
  pages   = {159--182},
  doi     = {10.1287/mnsc.14.3.159}
}

@book{aumann1995repeated,
  author    = {Aumann, Robert J. and Maschler, Michael B.},
  title     = {Repeated Games with Incomplete Information},
  publisher = {MIT Press},
  address   = {Cambridge, MA, USA},
  year      = {1995},
  series    = {MIT Press Classics}
}

@article{Hila2025,
  author  = {Hila, Angjelin},
  title   = {The epistemological consequences of large language models: Rethinking collective intelligence and institutional knowledge},
  journal = {AI \& Society},
  year    = {2025},
  doi     = {10.1007/s00146-025-02426-3},
}

@article{huang2024influence,
  title     = {The Influence of AI Literacy on User's Trust in AI in Practical Scenarios: A Digital Divide Pilot Study},
  author    = {Huang, Kuo Ting and Ball, Christopher},
  journal   = {Proceedings of the Association for Information Science and Technology},
  year      = {2024},
  volume    = {61},
  number    = {1},
  pages     = {937--939},
  doi       = {10.1002/pra2.1146},
  url       = {https://experts.illinois.edu/en/publications/the-influence-of-ai-literacy-on-users-trust-in-ai-in-practical-sc/}
}

@misc{ji2023ai,
  title        = {AI Alignment: A Comprehensive Survey},
  author       = {Jiaming Ji and Tianyi Qiu and Boyuan Chen and Borong Zhang and Hantao Lou and Kaile Wang and Yawen Duan and Zhonghao He and Jiayi Zhou and
                  Zhaowei Zhang and Fanzhi Zeng and Kwan Yee Ng and Juntao Dai and Xuehai Pan and Aidan O'Gara and Yingshan Lei and Hua Xu and Brian Tse and Jie Fu and
                  Stephen McAleer and Yaodong Yang and Yizhou Wang and Song-Chun Zhu and Yike Guo and Wen Gao},
  year         = {2023},
  eprint       = {2310.19852},
  archivePrefix= {arXiv},
  primaryClass = {cs.AI}
}

@article{ross1973agency,
  author  = {Ross, Stephen A.},
  title   = {The Economic Theory of Agency: The Principal's Problem},
  journal = {American Economic Review},
  year    = {1973},
  volume  = {63},
  number  = {2},
  pages   = {134--139}
}

@article{jensen1976theory,
  author  = {Jensen, Michael C. and Meckling, William H.},
  title   = {Theory of the Firm: Managerial Behavior, Agency Costs and Ownership Structure},
  journal = {Journal of Financial Economics},
  year    = {1976},
  volume  = {3},
  number  = {4},
  pages   = {305--360},
  doi     = {10.1016/0304-405X(76)90026-X}
}

@article{mirrlees1976optimal,
  author  = {Mirrlees, James A.},
  title   = {The Optimal Structure of Incentives and Authority Within an Organization},
  journal = {Bell Journal of Economics},
  year    = {1976},
  volume  = {7},
  number  = {1},
  pages   = {105--131},
  doi     = {10.2307/3003186}
}

@article{holmstrom1979moral,
  author  = {Holmstr\"om, Bengt},
  title   = {Moral Hazard and Observability},
  journal = {Bell Journal of Economics},
  year    = {1979},
  volume  = {10},
  number  = {1},
  pages   = {74--91},
  doi     = {10.2307/3003320}
}

@book{habermas2007reason,
  author = {Habermas, Jürgen and MacCarthy, Thomas and Habermas, Jürgen},
  title = {Reason and the Rationalization of Society},
  edition = {Nachdr.},
  series = {The Theory of Communicative Action / Jürgen Habermas. Transl. by Thomas MacCarthy, Vol. 1},
  address = {Boston},
  publisher = {Beacon},
  year = {2007}
}

@article{eisenstein2023helping,
  author    = {Jacob Eisenstein and Chirag Nagpal and Alekh Agarwal and Ahmad Beirami and Alex D'Amour and DJ Dvijotham and Adam Fisch and Katherine Heller and Stephen Pfohl and Deepak Ramachandran and Peter Shaw and Jonathan Berant},
  title     = {Helping or Herding? Reward Model Ensembles Mitigate but do not Eliminate Reward Hacking},
  journal   = {arXiv preprint arXiv:2312.09244},
  year      = {2023},
  doi       = {10.48550/arXiv.2312.09244},
  url       = {https://arxiv.org/abs/2312.09244}
}

@article{sharma2023towards,
  title     = {Towards Understanding Sycophancy in Language Models},
  author    = {Mrinank Sharma and Meg Tong and Tomasz Korbak and David Duvenaud and Amanda Askell and Samuel R. Bowman and Newton Cheng and Esin Durmus and Zac Hatfield-Dodds and Scott R. Johnston and Shauna Kravec and Timothy Maxwell and Sam McCandlish and Kamal Ndousse and Oliver Rausch and Nicholas Schiefer and Da Yan and Miranda Zhang and Ethan Perez},
  journal   = {arXiv preprint arXiv:2310.13548},
  year      = {2023},
  doi       = {10.48550/arXiv.2310.13548},
  url       = {https://arxiv.org/abs/2310.13548}
}

@inproceedings{chen2025selfaugmented,
  author    = {Chien Hung Chen and Hen-Hsen Huang and Hsin-Hsi Chen},
  title     = {Self-Augmented Preference Alignment for Sycophancy Reduction in LLMs},
  booktitle = {Proceedings of the 2025 Conference on Empirical Methods in Natural Language Processing},
  editor    = {Christos Christodoulopoulos and Tanmoy Chakraborty and Carolyn Rose and Violet Peng},
  pages     = {12379--12391},
  year      = {2025},
  address   = {Suzhou, China},
  publisher = {Association for Computational Linguistics},
  doi       = {10.18653/v1/2025.emnlp-main.625},
  url       = {https://aclanthology.org/2025.emnlp-main.625/}
}

@article{grossman1981informational,
  author  = {Grossman, Sanford J.},
  title   = {The Informational Role of Warranties and Private Disclosure about Product Quality},
  journal = {Journal of Law and Economics},
  year    = {1981},
  volume  = {24},
  number  = {3},
  pages   = {461--483},
  doi     = {10.1086/466995}
}

@article{milgrom1981good,
  author  = {Milgrom, Paul R.},
  title   = {Good News and Bad News: Representation Theorems and Applications},
  journal = {Bell Journal of Economics},
  year    = {1981},
  volume  = {12},
  number  = {2},
  pages   = {380--391},
  doi     = {10.2307/3003562}
}

@book{hardin1982collective,
  author    = {Hardin, Russell},
  title     = {Collective Action},
  publisher = {Johns Hopkins University Press},
  address   = {Baltimore},
  year      = {1982}
}

@article{arthur1983generalized,
  author  = {Arthur, W. Brian and Ermoliev, Yuri M. and Kaniovski, Yuri M.},
  title   = {On Generalized Urn Schemes of the P{\'o}lya Kind},
  journal = {Cybernetics},
  year    = {1983},
  volume  = {19},
  number  = {5},
  pages   = {61--71},
  issn    = {0011-4235},
  doi     = {10.1007/BF01068569}
}

@incollection{maynardsmith1972game,
  author    = {Maynard Smith, John},
  title     = {Game Theory and the Evolution of Fighting},
  booktitle = {On Evolution},
  publisher = {Edinburgh University Press},
  address   = {Edinburgh, UK},
  year      = {1972},
  pages     = {8--28}
}

@article{maynardsmith1973logic,
  author  = {Maynard Smith, John and Price, George R.},
  title   = {The Logic of Animal Conflict},
  journal = {Nature},
  year    = {1973},
  volume  = {246},
  number  = {5427},
  pages   = {15--18},
  doi     = {10.1038/246015a0}
}

@book{maynardsmith1982evolution,
  author    = {Maynard Smith, John},
  title     = {Evolution and the Theory of Games},
  publisher = {Cambridge University Press},
  address   = {Cambridge, UK},
  year      = {1982},
  isbn      = {9780521288843}
}

@book{skyrms1996evolution,
  author    = {Skyrms, Brian},
  title     = {The Evolution of the Social Contract},
  publisher = {Cambridge University Press},
  address   = {Cambridge, UK},
  year      = {1996},
  isbn      = {9780521555839}
}

@book{savage1954foundations,
  author    = {Savage, Leonard J.},
  title     = {The Foundations of Statistics},
  publisher = {John Wiley \& Sons},
  address   = {New York},
  year      = {1954},
  isbn      = {9780486623499}
}

@article{taylor1978evolutionary,
  author  = {Taylor, Peter D. and Jonker, Leo B.},
  title   = {Evolutionary Stable Strategies and Game Dynamics},
  journal = {Mathematical Biosciences},
  year    = {1978},
  volume  = {40},
  number  = {1-2},
  pages   = {145--156},
  doi     = {10.1016/0025-5564(78)90077-9}
}

@book{watson2013strategy,
  author    = {Watson, Joel},
  title     = {Strategy: An Introduction to Game Theory},
  edition   = {3rd},
  publisher = {W. W. Norton \& Company},
  address   = {New York},
  year      = {2013},
  isbn      = {9780393918380}
}

@article{fragiadakis2024evaluating,
  author  = {Fragiadakis, George and Diou, Christos and Kousiouris, George and Nikolaidou, Mara},
  title   = {Evaluating Human-AI Collaboration: A Review and Methodological Framework},
  journal = {arXiv preprint arXiv:2407.19098},
  year    = {2024},
  doi     = {10.48550/arXiv.2407.19098},
  url     = {https://arxiv.org/pdf/2407.19098}
}

@article{Cabreraetal2023,
  author  = {Cabrera, {\'A}ngel Alexander and Perer, Adam and Hong, Jason I.},
  title   = {Improving Human-AI Collaboration With Descriptions of AI Behavior},
  journal = {Proceedings of the ACM on Human-Computer Interaction},
  year    = {2023},
  volume  = {7},
  number  = {CSCW1},
  pages   = {1--21},
  doi     = {10.1145/3579612}
}

@article{Sarkar2023,
  author  = {Sarkar, Advait},
  title   = {Enough With "Human-AI Collaboration"},
  journal = {Extended Abstracts of the 2023 CHI Conference on Human Factors in Computing Systems (CHI EA '23)},
  year    = {2023},
  pages   = {1--8},
  doi     = {10.1145/3544549.3582735}
}

@article{Gomezetal2024,
  author  = {Gomez, Catalina and Cho, Sue Min and Ke, Shichang and Huang, Chien-Ming and Unberath, Mathias},
  title   = {Human--AI collaboration is not very collaborative yet: A taxonomy of interaction patterns in AI-assisted decision making from a systematic review},
  journal = {Frontiers in Computer Science},
  year    = {2024},
  volume  = {6},
  pages   = {1521066},
  doi     = {10.3389/fcomp.2024.1521066}
}

@article{PuertaBeldarrain2025,
  author  = {Puerta-Beldarrain, Maite and G{\'o}mez-Carmona, Oihane and S{\'a}nchez-Corcuera, Rub{\'e}n and Casado-Mansilla, Diego and L{\'o}pez-de-Ipi{\~n}a, Diego and Chen, Liming},
  journal = {IEEE Access}, 
  title   = {A Multifaceted Vision of the Human-AI Collaboration: A Comprehensive Review}, 
  year    = {2025},
  volume  = {13},
  pages   = {21876--21903},
  doi     = {10.1109/ACCESS.2025.3536095}
}

@misc{Fragiadakis2024,
  author        = {Fragiadakis, George and Diou, Christos and Kousiouris, George and Nikolaidou, Mara},
  title         = {Evaluating Human-AI Collaboration: A Review and Methodological Framework},
  year          = {2024},
  eprint        = {2407.19098},
  archivePrefix = {arXiv},
  primaryClass  = {cs.HC},
  doi           = {10.48550/arXiv.2407.19098}
}

@article{Zhu2025,
  author  = {Zhu, Li},
  title   = {How Does Collective Self-esteem Influence Confrontational Behaviors and Emotional Responses in Human-AI Competition?},
  journal = {Computers in Human Behavior: Artificial Humans},
  year    = {2025},
  volume  = {3},
  number  = {1},
  pages   = {100155},
  doi     = {10.1016/j.chbah.2025.100155},
  url     = {https://doi.org/10.1016/j.chbah.2025.100155}
}

@article{Noller2025,
  author  = {Noller, J{\"o}rg},
  title   = {4E cognition and the coevolution of human--AI interaction},
  journal = {Discover Artificial Intelligence},
  year    = {2025},
  volume  = {5},
  number  = {1},
  pages   = {323},
  doi     = {10.1007/s44163-025-00595-0},
  url     = {https://doi.org/10.1007/s44163-025-00595-0}
}

@incollection{Shkurko2025,
  author    = {Shkurko, Yulia S.},
  title     = {Human-AI Interaction from an Evolutionary Neurosociological Perspective},
  booktitle = {Handbook of Neurosociology},
  editor    = {Franks, David D. and Turner, Jonathan H. and Firat, Rengin B. and TenHouten, Warren D.},
  year      = {2025},
  pages     = {387--409},
  publisher = {Springer Nature Switzerland},
  series    = {Handbooks of Sociology and Social Research},
  doi       = {10.1007/978-3-031-95615-7_21},
  url       = {https://doi.org/10.1007/978-3-031-95615-7_21}
}

@phdthesis{Alsobay2025,
  author  = {Alsobay, Mohammed},
  title   = {Toward an Integrative Study of Human-AI Interaction},
  school  = {Massachusetts Institute of Technology},
  year    = {2025},
  address = {Cambridge, MA},
  month   = {September},
  note    = {Advisor: Abdullah Almaatouq},
  url     = {https://hdl.handle.net/1721.1/164569}
}

@article{Peeters2021,
  author  = {Peeters, Marieke M. M. and van Diggelen, Jurriaan and van den Bosch, Karel and Bronkhorst, Adelbert and Neerincx, Mark A. and Schraagen, Jan Maarten and Raaijmakers, Stephan},
  title   = {Hybrid collective intelligence in a human--AI society},
  journal = {AI \& SOCIETY},
  year    = {2021},
  volume  = {36},
  number  = {1},
  pages   = {217--238},
  doi     = {10.1007/s00146-020-01005-y},
  url     = {https://doi.org/10.1007/s00146-020-01005-y}
}

@article{Earp2025,
  author  = {Earp, Brian D. and Feroz, Faisal and Porsdam Mann, Sebastian and Voinea, Cristina and Chalson, Shalom and Jurcys, Paul and Reinecke, Madeline G. and Savulescu, Julian and Singh, Ilina and Clark, Margaret S.},
  title   = {How the Risk of Exploitation in Human-AI Relationships Depends on Relationship Type},
  journal = {SSRN Electronic Journal},
  year    = {2025},
  doi     = {10.2139/ssrn.5288102},
  url     = {https://ssrn.com/abstract=5288102}
}

@inproceedings{Wei2024,
  author    = {Wei, Zixi and Cao, Yuzhou and Feng, Lei},
  title     = {Exploiting Human-AI Dependence for Learning to Defer},
  booktitle = {Proceedings of the 41st International Conference on Machine Learning},
  series    = {Proceedings of Machine Learning Research},
  volume    = {235},
  pages     = {52201--52218},
  year      = {2024},
  publisher = {PMLR},
  url       = {https://openreview.net/forum?id=9H7WzF6qjK}
}

\end{document}